\documentclass[12pt]{article}
\usepackage[fleqn]{amsmath}
\usepackage{amssymb} 
\usepackage{hhline}
\usepackage[scale={.75,.75}]{geometry}
\usepackage[T1]{fontenc}
\usepackage{bbm} 
\usepackage{fleqn} 
\usepackage{graphicx} 
\usepackage{mathrsfs} 
\usepackage{cite} 
\usepackage{longtable}
\usepackage{caption}
\usepackage{axodraw}
\usepackage{cite}
\usepackage{epsfig}

\def\lsim{\mathrel{\raise.3ex\hbox{$<$\kern-.75em\lower1ex\hbox{$\sim$}}}}
\def\gsim{\mathrel{\raise.3ex\hbox{$>$\kern-.75em\lower1ex\hbox{$\sim$}}}}

\numberwithin{equation}{section}
\numberwithin{table}{section}
\allowdisplaybreaks

\begin{document}
\date{\mbox{ }}

\title{ 
{\normalsize     
DESY 08-127\hfill\mbox{}\\
September 2008\hfill\mbox{}\\}
\vspace{2cm}
{\Large\bf Superparticle Mass Window from Leptogenesis\\ 
and Decaying Gravitino Dark Matter}\\[4mm]}
%
\author{Wilfried Buchm\"{u}ller, Motoi Endo and Tetsuo Shindou \\[2mm]
{\normalsize\it Deutsches Elektronen-Synchrotron DESY, Hamburg, Germany}
}
\maketitle

\thispagestyle{empty}

\begin{abstract}
\noindent
Gravitino dark matter, together with thermal leptogenesis, implies an upper
bound on the masses of superparticles. In the case of broken R-parity the 
constraints from primordial nucleosynthesis are naturally satisfied and
decaying gravitinos lead to characteristic signatures in high energy cosmic
rays. We analyse the implications for supergravity models with 
universal boundary conditions at the grand unification scale. Together with 
low-energy observables one obtains a window of superparticle masses, which 
will soon be probed at the LHC, and a range of allowed reheating temperatures.
\end{abstract}

\newpage

\section{Introduction}

Standard thermal leptogenesis \cite{Fukugita:1986hr} provides a simple and
elegant explanation of the origin of matter. It is a natural consequence of
the seesaw mechanism, and it is perfectly consistent with the small neutrino
masses inferred from neutrino oscillation data \cite{Buchmuller:2005eh}.

Thermal leptogenesis works without and with supersymmetry. In the latter case,
however, there is a clash with the `gravitino problem' \cite{Weinberg:1982zq,
Ellis:1984er,Kawasaki:2004yh}: the large temperature required by leptogenesis
exceeds the upper bound on the 
reheating temperature from primordial nucleosynthesis 
(BBN) in typical supergravity models with a neutralino as lightest 
superparticle (LSP) and an unstable gravitino.   
If the gravitino is the LSP, the condition that relic gravitinos do not
overclose the universe yields an upper bound on the reheating temperature
\cite{Moroi:1993mb}. Furthermore, the next-to-lightest superparticle (NLSP) 
is long lived, and one has to worry about the effect of NLSP decays on 
nucleosynthesis.  

It is remarkable that, despite these potential problems, a large leptogenesis 
temperature of order $10^{10}$~GeV can account for
the observed cold dark matter in terms of thermally produced relic gravitinos
\cite{Bolz:1998ek}. Requiring consistency with nucleosynthesis  yields 
constraints on the superparticle mass spectrum. Due to improved analyses
of BBN, the original proposal of a higgsino NLSP is no longer viable,
and also other possible NLSPs are strongly constrained.   
The case of a stau NLSP is cornered by bounds following from 
catalyzed production of $^6{\rm Li}$ \cite{Pospelov:2006sc}, with
the possible exception of a large left-right mixing in the stau sector
\cite{Ratz:2008qh}. In some models a sneutrino \cite{Kanzaki:2006hm} or a stop 
\cite{DiazCruz:2007fc} can still be a viable NLSP. 

Recently, it has been shown that in the case of small R-parity and 
lepton number breaking, such that the baryon asymmetry is not erased
by sphaleron processes \cite{Campbell:1990fa}, thermal leptogenesis, gravitino
dark matter and primordial nucleosynthesis are naturally consistent 
\cite{Buchmuller:2007ui}. Although the gravitino is no longer stable, its
decay into standard model (SM) particles is doubly suppressed by the
Planck mass and the small R-parity breaking parameter. Hence, its lifetime
exceeds the age of the universe by many orders of magnitude, and it remains
a viable dark matter candidate \cite{Takayama:2000uz}. Gravitino decays lead
to characteristic signatures in high energy cosmic rays. The produced flux
of gamma-rays \cite{Takayama:2000uz,Buchmuller:2007ui,Bertone:2007aw,
Ibarra:2007wg,
Ishiwata:2008cu} and positrons \cite{Ibarra:2008qg,Ishiwata:2008cu} may 
explain the observed excess in the EGRET \cite{Strong:2004ry} and HEAT
\cite{Barwick:1997ig} data. This hypothesis will soon be tested by the
satellite experiments FGST and PAMELA.

In this paper we study the implications of leptogenesis and gravitino
dark matter with broken R-parity on the mass spectrum of superparticles.
Since the unification of gauge couplings in the minimal supersymmetric 
extension of the standard model (MSSM) is one of the main motivations   
for low-energy supersymmetry, we shall focus on versions of the MSSM
with universal boundary conditions for scalar and gaugino masses at the
grand unification (GUT) scale. As we shall see, the corresponding spectrum 
of superparticle masses will be fully covered at the LHC. This is the
main result of our analysis.

After some comments on R-parity violation in Section~2, we discuss the
lower bound on the reheating temperature from leptogenesis and the 
upper bound on the NLSP mass from gravitino dark matter in Section~3.
Section~4 deals with constraints on MSSM parameters from low-energy
observables, and the results of our numerical analysis are presented
in Section~5, followed by some conlusions in Section~6.

\section{Constraints on R-parity violation}
\label{Sec:R-parity}

Phenomenological aspects of R-parity violation have been widely discussed
in the literature \cite{Barbier:2004ez}. Here we are interested in the case 
of small R-parity
and lepton number breaking which was investigated in 
\cite{Takayama:2000uz,Buchmuller:2007ui,Ishiwata:2008cu}. The details
strongly depend on the flavour structure of R-parity violating couplings
and the pattern of supersymmetry breaking.
For completeness, we recall in the following the order of magnitude of
bounds on R-parity violating couplings, the corresponding lifetimes of
gravitino and NLSP, and in particular the dependence on the gravitino mass.

Stringent constraints on the lepton number and R-parity violating interactions
\begin{equation}\label{deltaL}
W_{\Delta L=1} = \lambda_{ikj} l_i e^c_j l_k + \lambda'_{kji} d^c_i q_j l_k
\end{equation}
are imposed by baryogenesis. Both operators contain lepton doublets.
Together with sphaleron processes they therefore influence the baryon
asymmetry at high temperature in the early universe. The requirement that 
an existing baryon asymmetry is not erased before the electroweak transition 
typically implies \cite{Campbell:1990fa}
\begin{equation}\label{rpv}
\lambda\ ,\lambda' < 10^{-7}\;.
\end{equation}
Remarkably, for such a small breaking of R-parity a gravitino
LSP has a lifetime much longer than the age of the universe 
\cite{Takayama:2000uz}
because of the double suppression of the decay rate by the inverse Planck 
mass and the R-parity breaking coupling. 
One then obtains for the gravitino lifetime (cf. \cite{Buchmuller:2007ui})
\begin{equation}\label{life}
\tau_{3/2}\ \sim \ 10^{25} {\rm s}\  
\left(\frac{\lambda}{10^{-8}}\right)^{-2}
\eta\left(\frac{\widetilde{m}}{m_{3/2}}\right)
\left(\frac{m_{3/2}}{100~{\rm GeV}}\right)^{-3}\;,
\end{equation}
where $\widetilde{m} \sim {\cal O}(100~\text{GeV})$ is a characteristic 
supersymmetry breaking mass scale. In the case of light gravitinos, 
$m_{3/2} \ll \widetilde{m}$, where only the decay into photon 
neutrino pairs is kinematically allowed, $\eta =1$ has been assumed in 
\cite{Buchmuller:2007ui}. For heavier gravitinos, decays into W-boson lepton 
and Z-boson lepton pairs are also possible, and we only know that
$\eta = {\cal O}(1)$ \cite{Ishiwata:2008cu}. In particular, the relation 
between gravitino lifetime and gravitino mass depends on the pattern of 
supersymmetry breaking.

In the case of a small breaking of R-parity, with an unstable gravitino LSP,
the NLSP lifetime becomes very short,
\begin{equation}
c\tau_{\rm NLSP} \sim 10~{\rm cm} \left({\lambda\over 10^{-8}}\right)^{-2}
\left(m_{\rm NLSP}\over {100~{\rm GeV}}\right)^{-1}\;.
\end{equation}
For couplings $\lambda, \lambda' > 10^{-14}$, the NLSP lifetime becomes
shorter than $10^3~\text{s}$. In case of a stau NLSP, superparticle decays 
then do not affect
the primordial abundances of light elements. Hence, baryogenesis,
primordial nucleosynthesis and gravitino dark matter can be consistent
in the range  
\begin{equation}
10^{-14} < \lambda,\lambda' < 10^{-7}\,.
\end{equation}
For a bino NLSP, a lifetime shorter than $0.1$~s, i.e., couplings 
$\lambda, \lambda' > 10^{-12}$ are required by consistency with BBN.

The analysis of
constraints on the superpotential terms (\ref{deltaL}) can be extended
to general R-parity breaking mass terms \cite{Ishiwata:2008cu}, yielding
again a range of allowed parameters. One finds that possible contributions
to neutrino masses are negligable, once the cosmological constraints are
satisfied. 

Decaying gravitino dark matter can contribute to the EGRET and HEAT
anomalies for a gravitino lifetime $\tau_{3/2} \sim 10^{26}$~s. For a
gravitino mass $m_{3/2} \sim 10$~GeV, and assuming $\eta \simeq 1$ in
Eq.~(\ref{life}), this requires R-parity violating
couplings $\lambda \sim 10^{-7}$. As we shall see, universal   
boundary conditions for gaugino masses favour larger gravitino masses,
in the range 
\begin{equation}
10~{\rm GeV} < m_{3/2} < 500~{\rm GeV}\;, 
\end{equation}
which, for fixed gravitino lifetime and $\eta \sim 1$, corresponds to the 
range of R-parity violating couplings
\begin{equation} 
10^{-10} < \lambda < 10^{-7}\;.
\end{equation} 
Note that for couplings below $\sim 10^{-9}$, most NLSPs decay outside 
the detector. However, for couplings above  $\sim 10^{-11}$, corresponding
to lifetimes shorter than  $\sim 10^{-3}~\text{s}$, some NLSP decays may still
be observable in the detector \cite{Ishiwata:2008tp}.  

How can the phenomenologically required small R-parity violating couplings
arise? In \cite{Buchmuller:2007ui} an example was presented, where the 
spontaneous breaking of R-parity is tied to B-L breaking. Recently, it
has been shown that also the breaking of left-right symmetry can lead to
small R-parity breaking \cite{Ji:2008cq}. 

\section{Thermal leptogenesis}
\label{Sec:ThermalLeptogenesis}

Let us now consider standard thermal leptogenesis as the source of the 
cosmological baryon asymmetry. In the high-temperature phase of the early
universe thermally produced right-handed neutrinos generate an asymmetry 
in B-L, which leads to a baryon asymmetry via sphaleron processes. 
In the case of hierarchical right-handed neutrinos, and neglecting
flavour effects, the baryon density relative to the photon density is 
given by (cf.~\cite{Buchmuller:2005eh})
\begin{equation}
\frac{n_B}{n_{\gamma}}
\;\simeq\; 
-1.04\times 10^{-2}\epsilon_1\kappa ,
\end{equation}
where $\epsilon_1$ is the CP asymmetry in the decay of the lightest 
right-handed neutrino $N_1$ into a pair of lepton ($L$) and Higgs ($H_u$) 
doublets, the efficiency factor $\kappa$ represents the effects of washout 
and scattering processes, and we have assumed a supersymmetric thermal plasma.
The CP asymmetry $\epsilon_1$ satisfies an upper bound because of the seesaw 
relation, which for supersymmetric models reads 
\cite{Hamaguchi:2001gw,Davidson:2002qv,Covi:1996wh},
\begin{eqnarray}
|\epsilon_1|
\;\equiv\;
\left|
\frac{\Gamma(N_1\to L+H_u)-\Gamma(N_1\to L^c+H_u^c)}
{\Gamma(N_1\to L+H_u)+\Gamma(N_1\to L^c+H_u^c)}
\right|
\;\lesssim\;
\frac{3M_1}{8\pi\langle H_u\rangle^2}
\frac{\Delta m_{\rm atm}^2}{m_1+m_3} .
\label{eq:maxeps1}
\end{eqnarray}
Here $m_i$, with $m_1 < m_2 < m_3$, are the mass eigenvalues of the light 
neutrinos and $M_1$ is the mass of the right-handed neutrino $N_1$.
The atmospheric neutrino mass squared difference is determined from
neutrino oscillation experiments as 
$\Delta m_{\rm atm}^2 \simeq (2.5 \pm 0.2) \times 10^{-3}\text{eV}^2$. 
Note that the upper bound on $|\epsilon_1|$, and therefore the maximally
generated baryon asymmetry, increases proportional to the heavy Majorana
mass $M_1$.

The efficiency factor $\kappa$ has to be determined by solving the Boltzmann 
equations. In the most interesting case of zero initial abundance of the 
right-handed neutrinos one finds for its maximal value, with and without
supersymmetry, $\kappa \simeq 0.2$ \cite{Buchmuller:2004nz,Giudice:2003jh}. 
Using (\ref{eq:maxeps1}), one then obtains from the observed baryon asymmetry 
\cite{Hinshaw:2008kr},
\begin{equation}
\frac{n_B}{n_{\gamma}} = (6.21\pm 0.16)\times 10^{-10}\; ,
\end{equation}
the lower bound on the right-handed neutrino mass 
\begin{equation}
M_1 \gsim 1.4\times 10^{9}\ \text{GeV}\,
\left(\frac{\langle H_u\rangle}{174\text{GeV}}\right)^2 
\label{eq:bound-MR}
\end{equation}
at the 3$\sigma$ level of $n_B/n_{\gamma}$ and $\Delta m_{\rm atm}^2$. 
The corresponding lower bound on the reheating temperature is about a factor
two smaller \cite{Blanchet:2006be}. In the following analysis we shall 
therefore use as an estimate
\begin{equation}
  T_R \;\gsim\; 1 \times 10^9\ {\rm GeV} .
  \label{eq:bound-TR}
\end{equation}
Note that this bound on the reheating temperature only applies for hierarchical
right-handed neutrinos. In the case of quasi-degenerate heavy neutrinos it
is relaxed. The bound also assumes thermal equilibrium, and it is modified
once the reheating process is taking into account. For instance, in the
case of reheating by inflaton decays, the bound increases by about a factor of
two \cite{Giudice:2003jh}. 

Relic gravitinos with masses larger than 1~GeV contribute to cold dark
matter. In the following analysis we identify the thermally produced 
abundance $\Omega_{3/2}h^2$ with the 2$\sigma$ upper bound on the dark
matter abundance deduced from the CMB anisotropies. From the WMAP 5-year 
results one obtains \cite{Hinshaw:2008kr}, 
\begin{equation}\label{dark}
  \Omega_{3/2}h^2 \equiv \Omega_{\rm DM}h^2 \simeq 0.1223\;.
\end{equation}

The thermal production of gravitinos is dominated by QCD processes. To leading
order in the gauge coupling we find
\begin{equation}\label{eq:gravitino-abundance}
\Omega_{3/2}h^2\ \simeq\ 0.5\  
\left(\frac{100\ {\rm GeV}}{m_{3/2}}\right)
\left(\frac{m_{\rm gluino}}{1\ {\rm TeV}}\right)^2
\left(\frac{T_R}{10^{10}\ {\rm GeV}}\right)\;,
\end{equation}
where $m_{\rm gluino}$ is the physical gluino mass. Note that
the coefficient\footnote{Varying superparticle masses (cf.~Section~4), the
value can change by about 10\%.} is about a factor two larger than in the 
analysis  
\cite{Bolz:2000fu}. This is due to the 2-loop running of 
the gluino mass, which has
been taken into account. Electroweak contributions to thermal gravitino
production further increase the abundance by about 20\%. In our numerical
analysis we shall take this into account following \cite{Pradler:2006qh}.
Note that the gravitino production rate has an ${\cal O}(1)$ uncertainty
due to unknown higher order contributions and nonperturbative effects
\cite{Bolz:2000fu}. Resummation of thermal masses increases the production
rate by about a factor of two \cite{Rychkov:2007uq}. We also neglect 
nonthermal contributions to gravitino production, in particular from
inflaton decay \cite{Endo:2007sz}, which are usually subdominant at
the considered high temperatures.

Our main interest are constraints on gluino and NLSP masses for gravitino
dark matter. It is then convenient to rewrite (\ref{eq:gravitino-abundance})
as
\begin{equation}\label{NLSP}
m_{\rm NLSP}\ \simeq\ 310\ {\rm GeV}\ \left(\frac{\xi}{0.2}\right)
\left(\frac{m_{3/2}}{100\ {\rm GeV}}\right)^{1/2} 
\left(\frac{10^9\ {\rm GeV}}{T_R}\right)^{1/2} , \quad
\xi = \frac{m_{\rm NLSP}}{m_{\rm gluino}}\ ,
\end{equation}
where the ratio $\xi$
is fixed by the boundary conditions of the soft supersymmetry breaking 
parameters.
For each gravitino mass and reheating temperature, Eq.~(\ref{NLSP})
then gives the NLSP mass for which the observed dark matter density is 
obtained. The maximal NLSP mass is reached for $m_{3/2} = m_{\rm NLSP}$,
\begin{equation}\label{NLSPmax}
m_{\rm NLSP}\ \lsim\ 980\ {\rm GeV}\ \left(\frac{\xi}{0.2}\right)^2
\left(\frac{10^9\ {\rm GeV}}{T_R}\right)\;.
\end{equation}

In this paper, we focus on thermally produced gravitino dark matter.
A high reheating temperature can also be consistent with leptogenesis 
in the case of very heavy gravitinos, as in anomaly mediation \cite{AMSB}
or mirage mediation \cite{CFNO,EYY}. In those models, the gravitino 
can have a mass of about 100~TeV and thus decays before BBN starts. 
However, these models have several intrinsic difficulties. In the case of 
anomaly mediation, it is difficult to explain the $g-2$ anomaly together
with the $b \to s\gamma$ constraint, since the gaugino masses are controlled 
by the beta functions. In mirage mediation models, one often has a light 
modulus field whose decay produces too many gravitinos \cite{NonthGrav}. 
Hence, the heavy gravitino scenario appears to be phenomenologically 
disfavoured.

\section{Models and low-energy observables}
\label{Sec:ModelsAndLowEnergy}

In order to illustrate the implications of leptogenesis and gravitino dark
matter on superparticle masses, we now study two typical boundary conditions
for the supersymmetry breaking parameters of the MSSM at the grand unification 
(GUT) scale:
\begin{eqnarray}
  (\text{A})~~~m_0 = m_{1/2},~~a_0 = 0,~~\tan\beta\;,
\end{eqnarray}
with equal universal scalar and gaugino masses, $m_0$ and $m_{1/2}$, 
respectively; in this case a bino-like neutralino becomes the NLSP. The second 
boundary condition is
\begin{eqnarray}
  (\text{B})~~~m_0 = 0,~~m_{1/2},~~a_0 = 0,~~\tan\beta\;,
\end{eqnarray}
which yields the right-handed stau as NLSP. In both cases, the trilinear 
scalar coupling $a_0$ is put to zero for simplicity. The ratio $\tan\beta$ 
of the Higgs vacuum expectation values and the universal gaugino mass 
$m_{1/2}$ are the two remaining independent variables. Superparticle masses
at the electroweak scale are obtained by solving the renormalization group 
equations at 2-loop accuracy by means of {\tt SOFTSUSY} 2.0.18
\cite{Allanach:2001kg}. 

Low-energy observables yield a lower bound on superparticle masses. Since
the thermal gravitino abundance (\ref{eq:gravitino-abundance}) increases
quadratically with the gluino mass, this implies an upper bound on the
reheating temperature. Together with the lower bound from leptogenesis
one then obtains a range of allowed reheating temperatures. In the same
way, leptogenesis and gravitino dark matter yield an upper bound on 
superparticle masses. Combined with low-energy constraints, a window of 
allowed superparticle masses is obtained.

One of the strongest constraints on the MSSM parameter space follows from the 
lower bound on the Higgs boson mass by LEP \cite{PDG},
\begin{eqnarray}
  m_h \;>\; 114.4\ {\rm GeV}~~~(95\%{\rm C.L.})\ .
\end{eqnarray}
The bound is satisfied by enhancing radiative corrections to the Higgs 
potential, which requires a large stop mass. The parameters of the stop sector
are essentially controlled by the gluino mass, i.e. $m_{1/2}$, via the 
renormalization group evolutions; they are less sensitive to the scalar mass 
$m_0$. The potential is also affected by the trilinear stop coupling 
$A_t$ for sufficiently large $a_0$. Although we put $a_0 = 0$ in the
numerical analysis, we shall comment on the case $a_0\neq 0$. In our analysis 
we use the top quark mass $m_t = 172.6$~GeV \cite{PDG}. Radiative corrections 
are taken into account at the 2-loop level by means of {\tt FeynHiggs} 2.6.4 
\cite{Frank:2006yh}. 

When the superparticles are light, they contribute significantly to rare  
processes. The measured branching ratio ${\rm Br}(B_d  \to X_s \gamma)$ 
agrees with the SM prediction. The SUSY contributions are dominated by the 
top-charged Higgs and stop-chargino diagrams. The latter is enhanced by 
large $\tan\beta$ and interferes with the former. In our analysis we choose
the sign of the supersymmetric Higgs mass parameter $\mu_H$ such that the 
effect of the SUSY contributions is reduced. Taking into account the 
theoretical uncertainties, we require for the full MSSM 
prediction the conservative upper and lower bounds, 
\begin{eqnarray}
  2 \times 10^{-4} \;<\; 
  {\rm Br}(B_d \to X_s \gamma) \;<\; 
  4 \times 10^{-4}.
\end{eqnarray}
The numerical analysis is based on {\tt SusyBSG} 1.1.2 which takes NNLO
contributions partially into account \cite{Degrassi:2007kj}.

The two observables discussed above constrain the MSSM parameters. In contrast,
the apparent discrepancy between the measured value of the muon anomalous 
magnetic moment \cite{Bennett:2006fi} and the SM prediction may be an effect of
supersymmetry, which then favours a certain range of MSSM parameters. Recently,
the hadronic contribution to the SM prediction has been updated using $e^+e^-$ 
data \cite{Hagiwara:2006jt}. The current discrepancy with experiment 
is given by \cite{Passera:2008jk}
\begin{eqnarray}
  a_\mu({\rm exp}) - a_\mu({\rm SM}) =
  302(88) \times 10^{-11}\;,
  \label{eq:muong-2}
\end{eqnarray}
which corresponds to a 3.4$\sigma$ deviation. An explanation of this
discrepancy by hypothetical errors in the determination of the hadronic SM
contribution appears unlikely \cite{Passera:2008jk}. In contrast,
supersymmetry can easily account for the discrepancy \cite{SUSYg-2}. The SUSY 
contribution is proportional to $\tan\beta$ and depends on ${\rm sgn}(\mu_H)$.
It is remarkable that the deviation from the SM prediciton for  $a_\mu$ and
the agreement for ${\rm Br}(B_d \to X_s \gamma)$ require the same  
${\rm sgn}(\mu_H)$ in the case of universal gaugino masses at the GUT scale. 
In the following, we use {\tt FeynHiggs} to evaluate the SUSY contribution to 
$a_\mu$ at the 2-loop level.

Finally, the absense of pair production of heavy charged particles at LEP 
implies the approximate lower mass bound \cite{PDG}
\begin{eqnarray}
  m_{\rm charged} > 100\ {\rm GeV}\;.
\end{eqnarray}
In the next section we shall use superparticle masses obtained by means of
{\tt SOFTSUSY}.

\section{Numerical analysis}
\label{Sec:Numerical}

We are now ready to determine the superparticle mass window and the allowed
range of reheating temperatures for the two examples of universal boundary
conditions at the GUT scale, which were discussed in the previous section.

\begin{figure}[t]
\begin{tabular}{cc}
\includegraphics[scale=0.6]{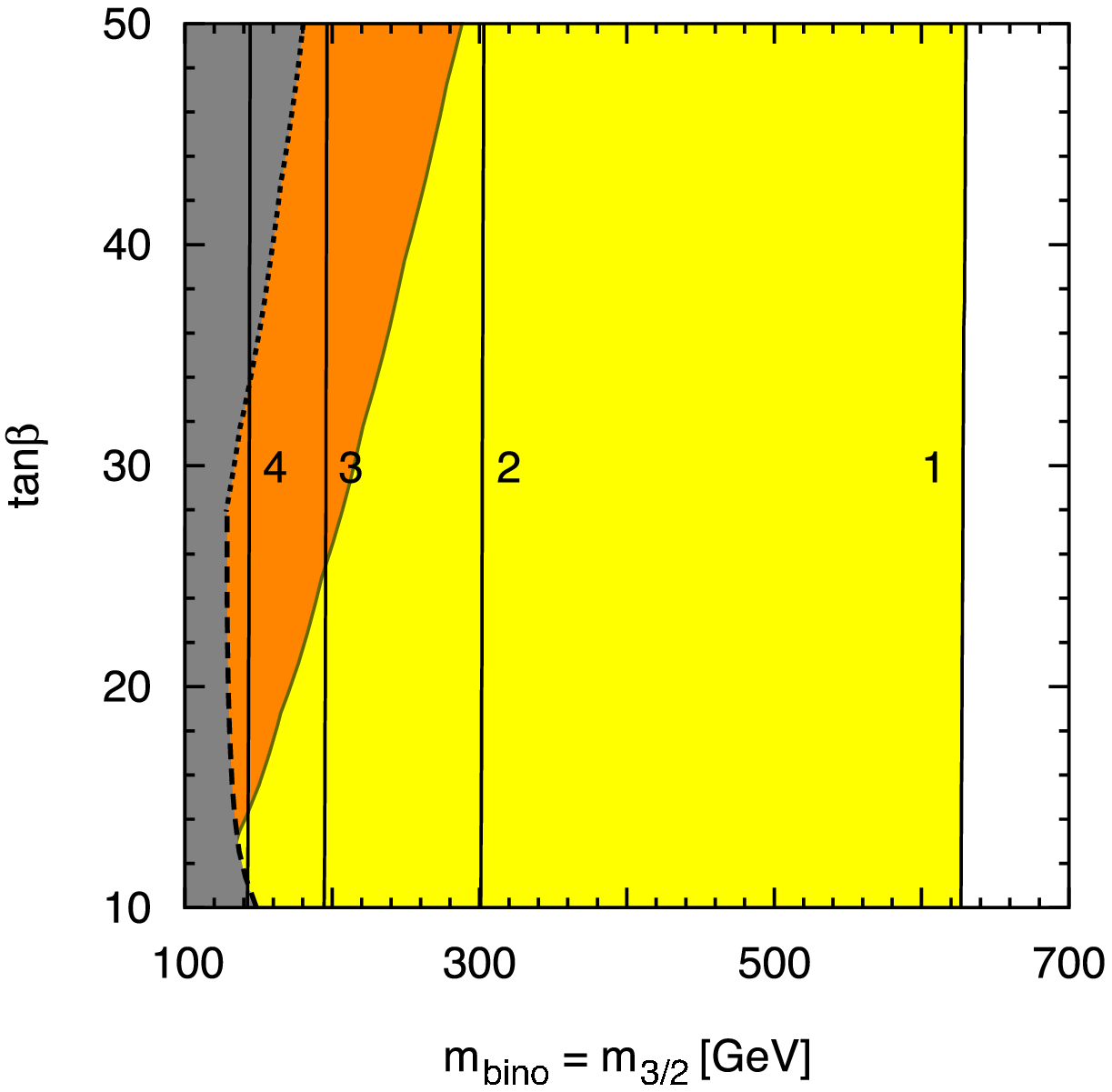} &
\includegraphics[scale=0.6]{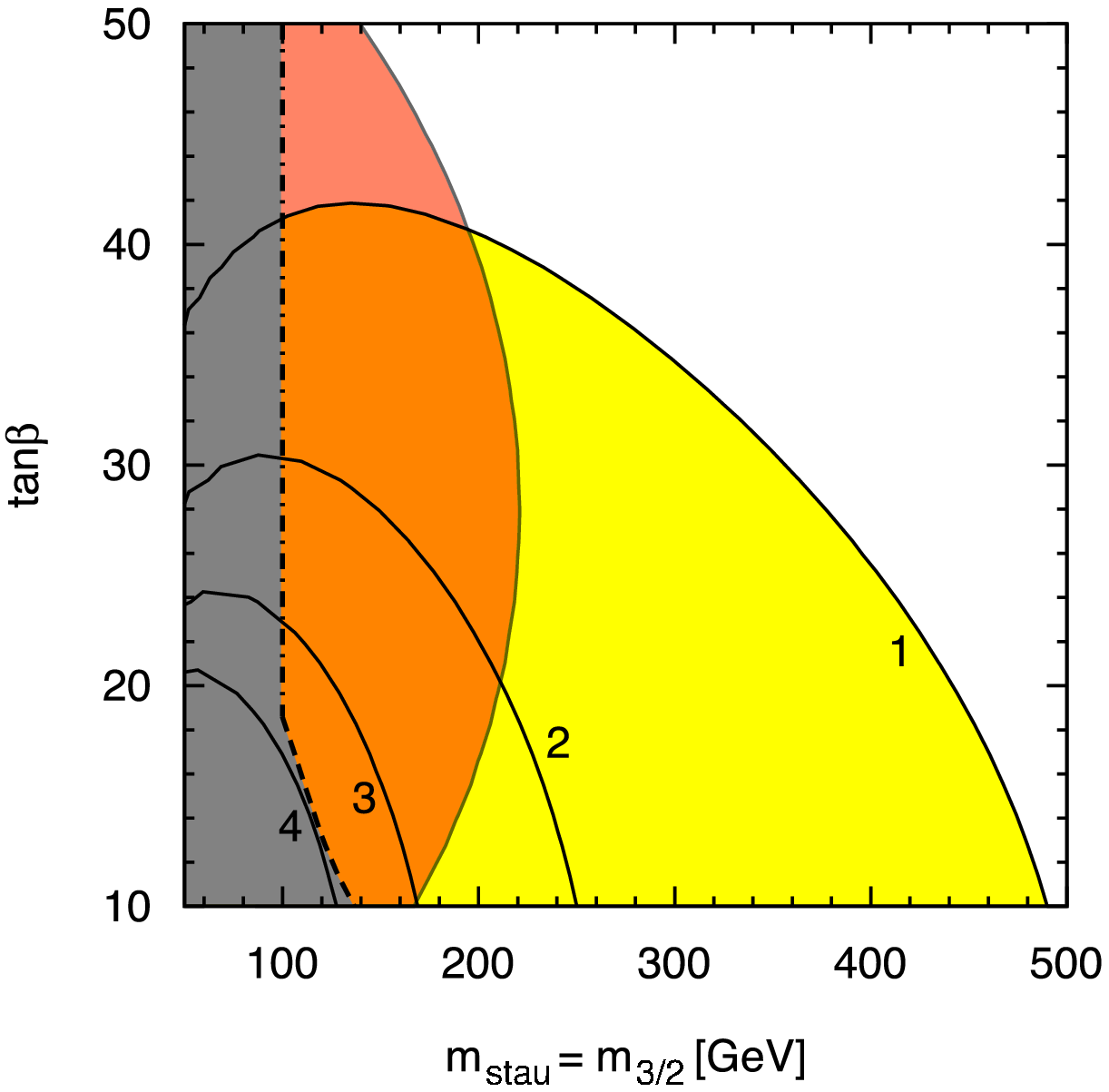} \\
(a) & (b)
\end{tabular}
\caption{
Contours of constant reheating temperature, $T_R = (1-4) \times 10^9$~GeV, 
with $\Omega_{3/2} = \Omega_\text{DM}$ (solid lines) (cf.~Eq.~(\ref{dark})). 
The panels (a) and
(b) correspond to the GUT boundary conditions (A) and (B) with bino-like
NLSP and stau NLSP, respectively. The choice $m_{3/2}=m_\text{NLSP}$
maximizes the reheating temperature. 
The gray region is excluded by constraints from low-energy 
experiments: the lower $\tan\beta$ part (left of the dashed line) does not 
satisfy the LEP Higgs mass bound; the higher $\tan\beta$ part in (a) (left
of the dotted line) is ruled out by ${\rm Br}(B_d \to X_s \gamma)$;
the higher $\tan\beta$ part in (b) (left of the dot-dashed line) does not
satisfy the lower mass bound on charged particles from LEP.
Thermal leptogenesis is possible in the yellow and orange regions; the
orange region is favored by the muon $g-2$ anomaly at the 2$\sigma$ level. 
}
\label{fig:TR}
\end{figure}

In Fig.~\ref{fig:TR} the upper bound (\ref{NLSPmax}) on the NLSP masses is 
shown for reheating temperatures $T_R \geq 1 \times 10^9$~GeV, which is the 
lower bound required by leptogenesis. In case (A) with bino NLSP, the ratio
$\xi = m_{\rm NLSP}/m_{\rm gluino}$, and therefore the upper bound on 
$m_{\rm NLSP}$, are essentially independent of $\tan \beta$. In contrast,
for (B) with stau NLSP, one has a strong dependence on $\tan \beta$.   
The lower bound on $m_{\rm NLSP}$ is determined by 
${\rm Br}(B_d \to X_s \gamma)$ and the Higgs mass bound in case (A), and the
charged particle and Higgs mass bounds in case (B), respectively. We find
the allowed mass ranges
\begin{equation}
(\text{A})~~130~{\rm GeV} < m_{\rm bino} < 620~{\rm GeV}\;, \quad
(\text{B})~~100~{\rm GeV} < m_{\rm stau} < 490~{\rm GeV}\;.
\end{equation}
Note that in case (B) upper and lower bounds correspond to different values
of $\tan \beta$. The muon g-2 anomaly favours small NLSP masses in the range
from $100~{\rm GeV}$ to $300~{\rm GeV}$. One also obtains upper bounds on the 
gravitino mass,
\begin{equation}
(\text{A})~~m_{3/2} < 620~{\rm GeV}\;, \qquad
(\text{B})~~m_{3/2} < 490~{\rm GeV}\;.
\end{equation}

Both boundary conditions have $a_0 = 0$. For negative $a_0$, 
the Higgs boson potential is modified in such a way that the dashed line 
in Fig.~\ref{fig:TR} moves to the left. We have checked that the reheating 
temperature can then reach $6 \times 10^9$~ GeV, whereas other observables are 
not much affected. We therefore obtain for the range of reheating temperatures
consistent with leptogenesis and gravitino dark matter
\begin{equation}
T_R = (1 - 6) \times 10^9~{\rm GeV}\;.
\end{equation} 
Note that according to {\tt FeynHiggs}, the theoretical uncertainty of the
Higgs boson is about $1~{\rm GeV}$ for $m_h \simeq 115$~GeV. This corresponds
to an uncertainty of  $10-20$\% for the upper bound on the reheating
temperature.

\begin{figure}[t]
\begin{tabular}{cc}
\includegraphics[scale=0.6]{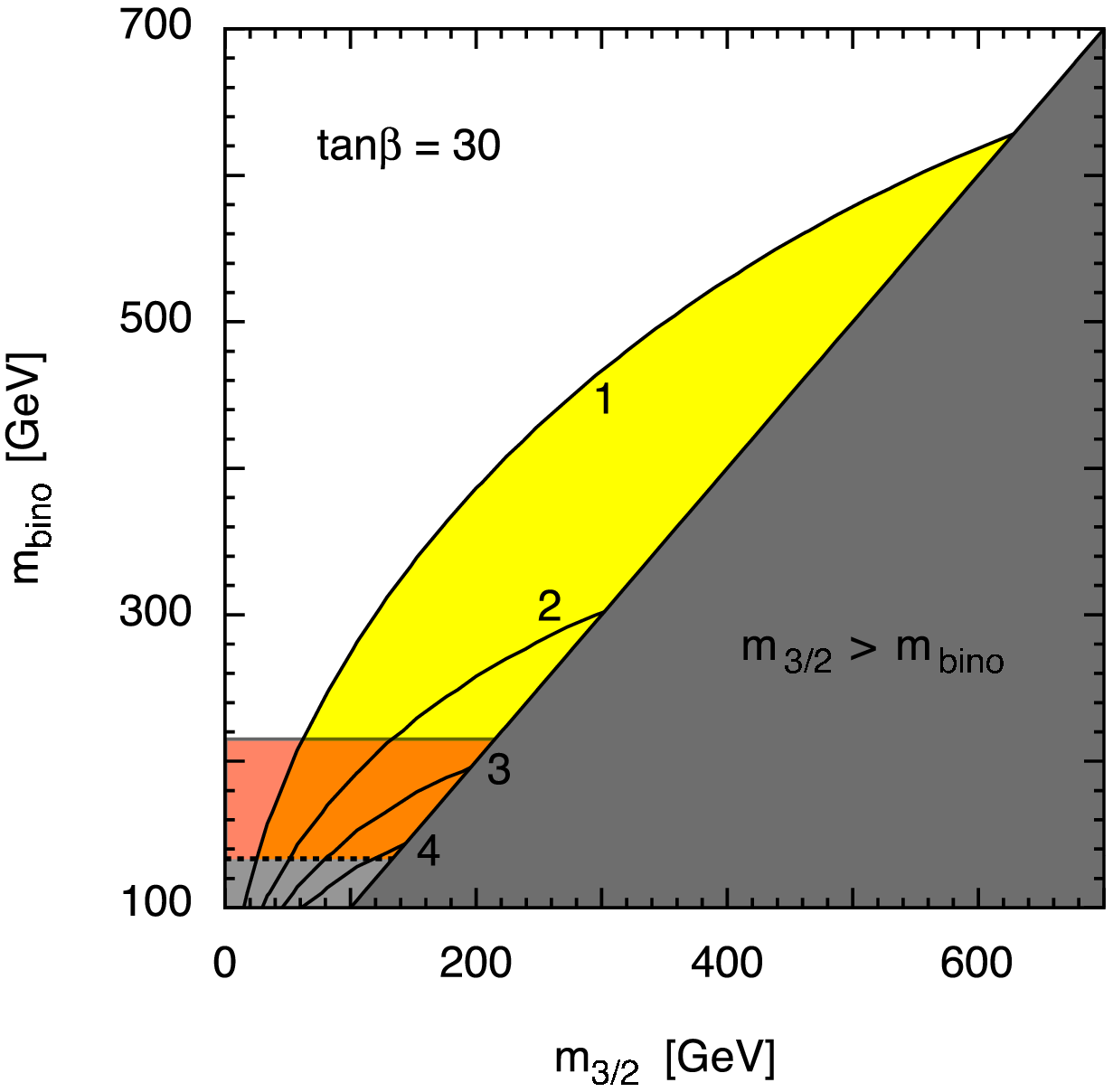}&
\includegraphics[scale=0.6]{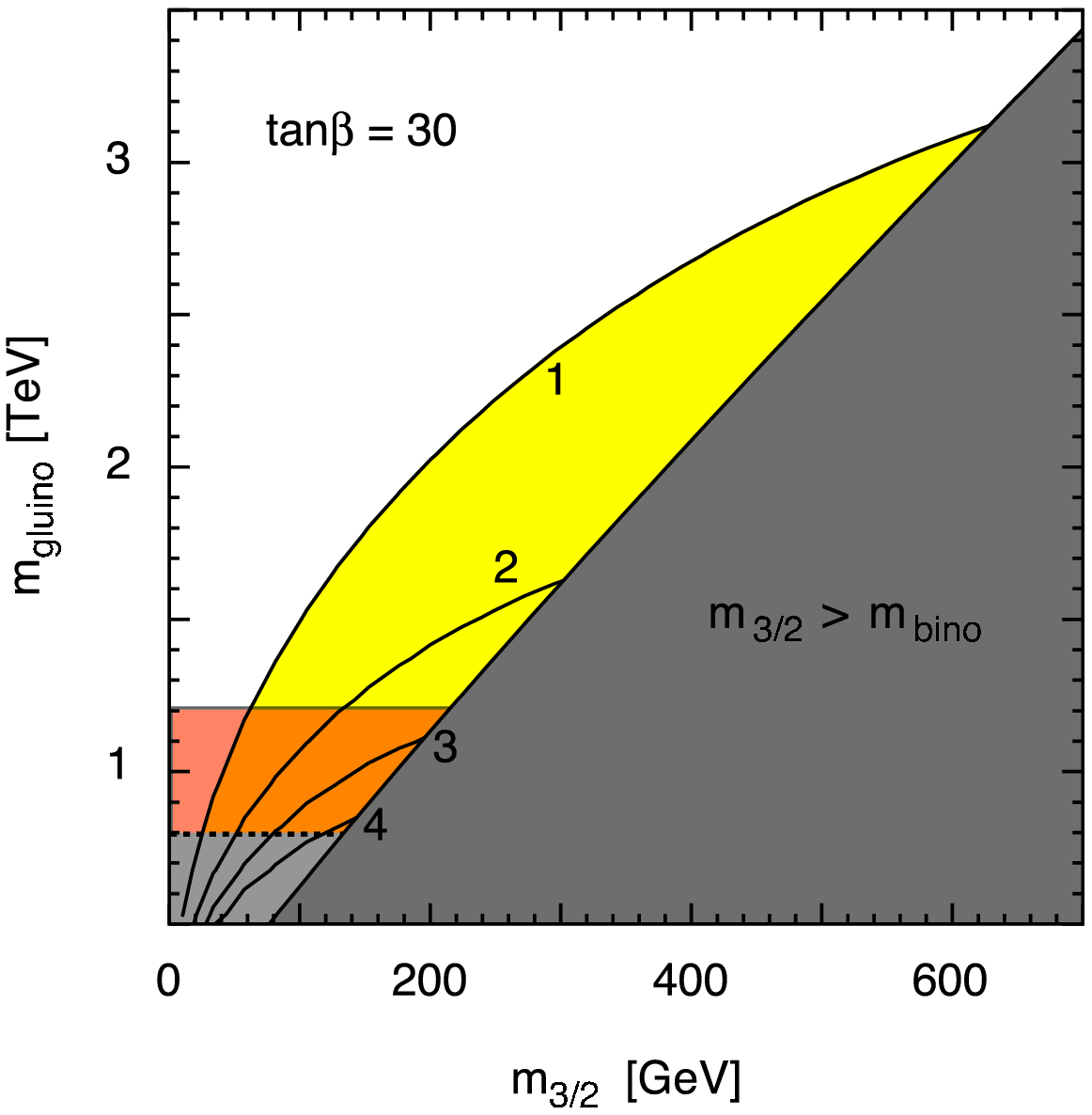}\\
(a)&(b)\\
\end{tabular}
\caption{
Contours of constant reheating temperature in the $m_\text{bino}-m_{3/2}$
plane (a) and the $m_\text{gluino}-m_{3/2}$ plane (b) for boundary condition
(A) with bino NLSP (see caption of Fig.~\ref{fig:TR} for details). In the 
dark gray region, the gravitino is not the LSP.
}
\label{fig:msugra}
\end{figure}

We can also study superparticle masses as function of gravitino mass and 
reheating temperature using Eq.~(\ref{NLSP}). The allowed NLSP mass range then
depends on $\tan \beta$. In the case of bino NLSP, consider as an example
\begin{equation}
(\text{A})~~\tan\beta = 30\;,\quad 
\xi = \frac{m_{\rm bino}}{m_{\rm gluino}} = 0.17 - 0.19\;.
\end{equation}
The left panel of Fig.~\ref{fig:msugra} shows the bino mass yielding the 
observed dark matter abundance as function of the gravitino mass for different
reheating temperatures; the right panel is the corresponding plot for the 
gluino mass. Upper mass bounds are obtained for the smallest temperature of 
$1\times 10^9$~GeV and the largest gravitino mass $m_{3/2}=m_{\rm bino}$,
\begin{equation}
(\text{A})~~\tan\beta = 30:\; m_{\rm bino} \lsim 620~{\rm GeV}\;,\quad 
m_{\rm gluino} \lsim 3.1~{\rm TeV}\;.
\end{equation}
For smaller gravitino masses the bounds become more stringent. For instance,
for $m_{3/2} = 100$~GeV, one obtains
\begin{equation}
m_{\rm bino} \lsim 270~{\rm GeV}\;, \quad 
m_{\rm gluino} \lsim 1.5~{\rm TeV}\;. 
\end{equation}
Note that these bounds are essentially independent of $m_0$ and $\tan\beta$,
als long as $m_0 \sim m_{3/2}$.

\begin{figure}[h]
\begin{tabular}{cc}
\includegraphics[scale=0.6]{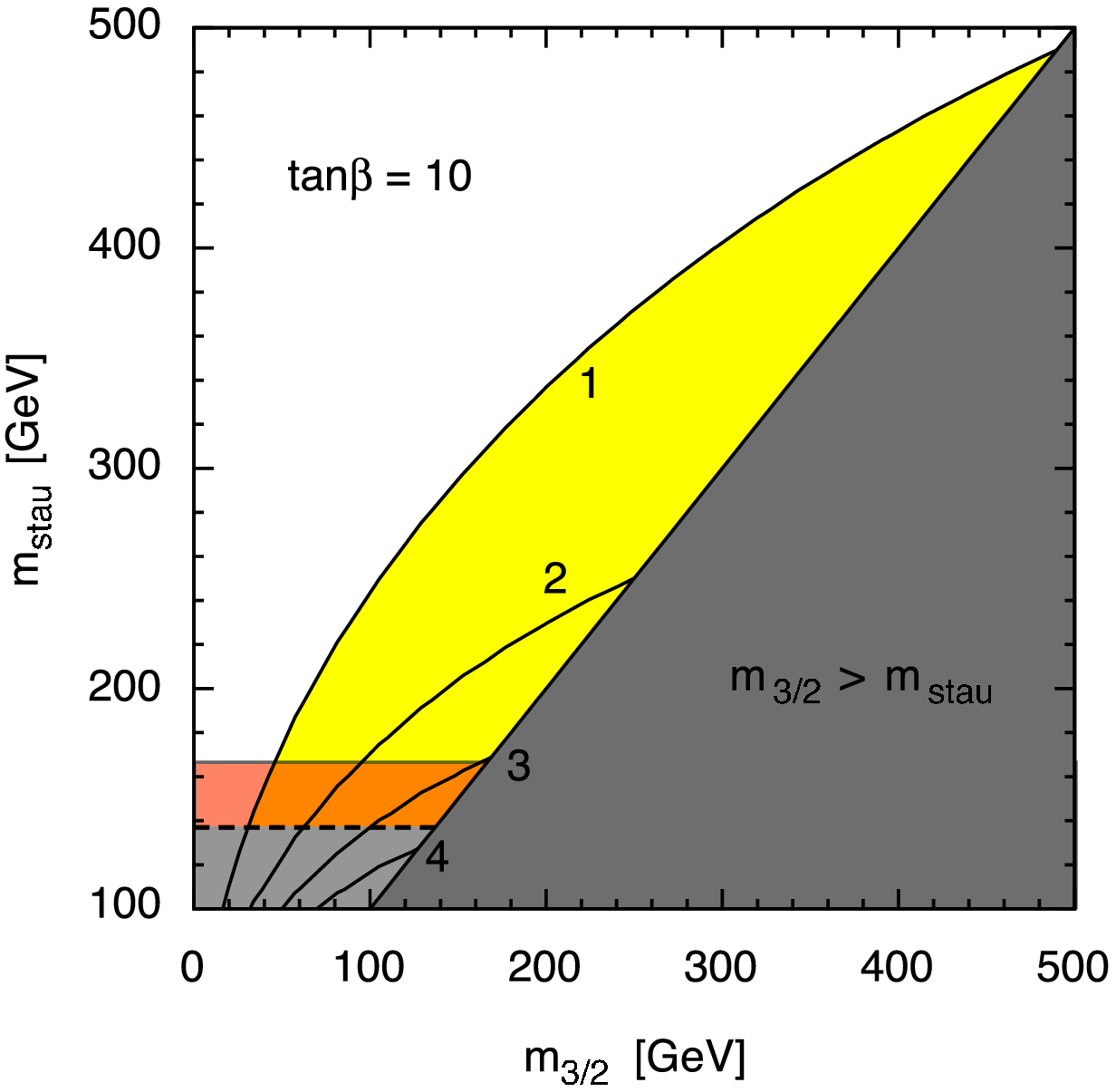}&
\includegraphics[scale=0.6]{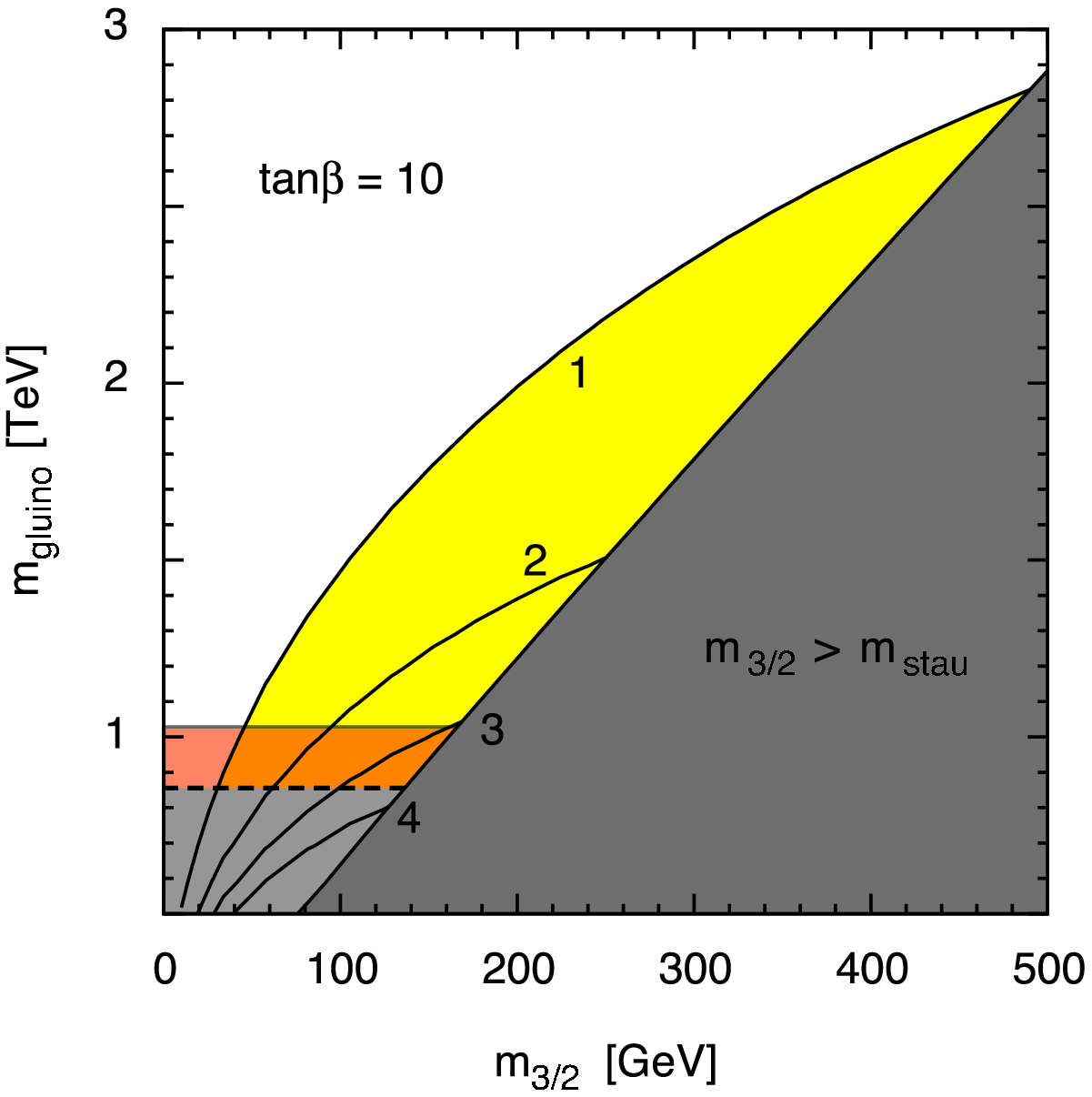}\\
(a)&(b)\\
\end{tabular}
\caption{
Contours of constant reheating temperature in the $m_\text{stau}-m_{3/2}$
plane (a) and the $m_\text{gluino}-m_{3/2}$ plane (b) for boundary condition
(B) with stau NLSP (see caption of Fig.~\ref{fig:TR} for details). In the 
dark gray region, the gravitino is not the LSP.
}
\label{fig:noscale}
\end{figure}

In the case of stau NLSP, there is a strong dependence on $\tan\beta$. As an
example, we consider
\begin{equation}
(\text{B})~~\tan\beta = 10\;,\quad 
\xi = \frac{m_{\rm stau}}{m_{\rm gluino}} = 0.16 - 0.17\;.
\end{equation}
$\xi$ decreases with increasing $\tan{\beta}$.
Stau and gluino masses are shown in Fig.~3. Since the ratio of NLSP and gluino
mass is smaller, the mass bounds are now more stringent, 
\begin{equation}
(\text{B})~~\tan\beta = 10:\; m_{\rm stau} \lsim 490~{\rm GeV}\;,\quad 
m_{\rm gluino} \lsim 2.8~{\rm TeV}\;.
\end{equation}
For a gravitino mass $m_{3/2} = 100$~GeV, one obtains
\begin{equation}
m_{\rm stau} \lsim 240~{\rm GeV}\;, \quad 
m_{\rm gluino} \lsim 1.5~{\rm TeV}\;. 
\end{equation}

Let us emphasize again the effect of the theoretical uncertainty in the 
evaluation of the gravitino abundance, which is expected to be 
${\cal O}(1)$ \cite{Bolz:2000fu}. For instance,
if the gravitino production rate is larger by a factor 2, 
as suggested in \cite{Rychkov:2007uq}, all reheating temperatures in 
Figs.~\ref{fig:TR}, \ref{fig:msugra} and \ref{fig:noscale} are by a 
factor 2 smaller. Hence, the superparticle mass range consistent
with thermal leptogenesis becomes narrower. On the other hand, a smaller
gravitino production rate would enlarge the parameter range consistent
with leptogenesis.

Finally, let us comment on other boundary conditions. 
We have chosen universal gaugino masses, with $m_0 = m_{1/2}$ or  
$m_0 =0$ at the GUT scale. However, even for non-universal gaugino masses 
we obtain almost the same results. The reason is that all the bounds are 
controlled by the gluino mass. Reducing the gluino mass, the dark matter
bound on the reheating temperature is relaxed, but the low-energy
constraints become severer: supersymmetric contributions to the Higgs boson
mass are suppressed, while they are enhanced for 
${\rm Br}(B_d \to X_s \gamma)$. As a consequence,
the maximal reheating temperature remains almost
the same as in the case of universal gaugino masses. On the other hand,
the low-energy 
constraints become weaker for scalar masses much larger than $m_{1/2}$. 
One can then reach reheating temperatures $\sim 10^{10}$GeV.

\section{Conclusions and outlook}
\label{Sec:Conclusion}

We have studied the implications of thermal leptogenesis and gravitino dark 
matter for the mass spectrum of superparticles. In the case of broken
R-parity the constraints from nucleosynthesis are naturally fulfilled, and
universal gaugino masses at the GUT scale are possible, contrary to the
case of stable gravitinos.

As an illustration, we have considered two boundary conditions which lead
to a bino-like NLSP and a stau NLSP, respectively. Low-energy observables and
gravitino dark matter together with thermal leptogenesis yield upper and lower
bounds on NLSP and gluino masses, which in both cases lie within the
discovery range of the LHC. It is encouraging that the supersymmetric
explanation of the muon $g-2$ anomaly favours smaller masses within these
mass windows.

A cosmology with leptogenesis and gravitino dark matter also leads to the
prediction of a maximal temperature in the early universe. In the case of 
universal gaugino masses at the unification scale we find the upper bound 
$T_R^{\rm max}\simeq 6\times 10^9$~GeV, which is somewhat relaxed for
large scalar masses. This bound has been obtained under the 
assumption of thermal equilibrium, which appears unlikely for a maximal
temperature. Nevertheless, it is intriguing that the temperature 
$T_R^{\rm max}$ is of the same order of magnitude as the critical for
the destabilization of compact dimensions in higher-dimensional
supersymmetric theories \cite{Buchmuller:2004xr}. The effect of the
reheating process on the stabilization of extra dimensions and the
relation to baryogenesis and dark matter require futher investigations.

Gravitino decays produce a flux of photons and positrons, which
can significantly contribute to the EGRET and HEAT anomalies for a lifetime 
$\tau_{3/2} \sim 10^{26}$~s. If these anomalies are indeed related to gravitino
decays, the satellite experiments FGST and PAMELA should soon detect
characteristic features in the photon and positron spectrum, respectively. 
Observation of a line in the gamma-ray spectrum by FGST and a rise with sharp
cutoff in the positron spectrum by PAMELA would lead to a determination of the 
gravitino mass. This would considerably tighten the predictions for 
superparticle mass windows which will be probed at the LHC.



\end{document}